\documentclass[conference]{IEEEtran}
\IEEEoverridecommandlockouts

\usepackage{cite}

\ifCLASSINFOpdf
\else
  \usepackage[dvips]{graphicx}
\fi
\usepackage[cmex10]{amsmath}
\usepackage{array}
\usepackage[tight,footnotesize]{subfigure}
\usepackage{url}
\usepackage{amsthm}
\usepackage{amssymb}
\usepackage{color}
\theoremstyle{plain}

\theoremstyle{remark}

\begin{document}
\title{Massive MIMO for Crowd Scenarios: A Solution Based on Random Access}

\author{
\IEEEauthorblockN{
Jesper H. S\o rensen, Elisabeth de Carvalho and Petar Popovski
}
\IEEEauthorblockA{
Aalborg University, Department of Electronic Systems,
Fredrik Bajers Vej 7, 9220 Aalborg, Denmark
\\
E-mail: \{jhs,edc,petarp\}@es.aau.dk
\thanks{The research presented in this paper was partially supported by the Danish Council for Independent Research (Det Frie Forskningsr{\aa}d) DFF - $1335-00273$. Part of this work has been performed in the framework of the FP7 project ICT-317669 METIS, which is partly funded by the European Union. The authors would like to acknowledge the contributions of their colleagues in METIS, although the views expressed are those of the authors and do not necessarily represent the project.}
}
}
\maketitle

\begin{abstract}
This paper presents a new approach to intra-cell pilot contamination in crowded massive MIMO scenarios. The approach relies on two essential properties of a massive MIMO system, namely near-orthogonality between user channels and near-stability of channel powers. Signal processing techniques that take advantage of these properties allow us to view a set of contaminated pilot signals as a graph code on which iterative belief propagation can be performed. This makes it possible to decontaminate pilot signals and increase the throughput of the system. The proposed solution exhibits high performance with large improvements over the conventional method. The improvements come at the price of an increased error rate, although this effect is shown to decrease significantly for increasing number of antennas at the base station.
\end{abstract}
\IEEEpeerreviewmaketitle

\section{Introduction} \label{sec:introduction}
Multiple-input multiple-output (MIMO) has been identified as a key technology to improve spectral efficiency of wireless communication systems and is finding its way into practical systems, like LTE and LTE-Advanced. The research in MIMO has recently took a turn, when the advantage of having a massive number of antennas at a base station (BS) was asserted in \cite{marzetta06}. In \cite{marzetta06}, a massive MIMO system refers to a multi-cell multi-user system with a massive number of antennas at the BS that serves multiple users. The number of users is much smaller than the number of BS antennas, defining an under-determined multi-user system with a massive number of extra spatial degrees of freedom (DoF). Exploiting those extra DoF and assuming an infinite number of antennas at the BS, the multi-user MIMO channel can be turned into an orthogonal channel and the effect of small-scale fading and thermal noise can be eliminated. Based on those excellent properties, massive MIMO is acknowledged as a promising technology for very high system throughput and energy efficiency \cite{Petar5G}. 

When the number of antennas becomes massive, acquiring the channel state information (CSI) becomes a severe bottleneck. Downlink channel training requires a training length that is proportional to the number of antennas at the BS and is thus impractical. One solution promoted in \cite{marzetta06} restricts massive MIMO operations to time-division duplex (TDD) for which channel reciprocity is exploited. As the downlink and uplink channels are equal, CSI is acquired at the BS based on uplink training and then used for downlink transmission. The benefit is that the training length is proportional to the number of users, which is much smaller than the number of BS antennas. 

As described in \cite{marzetta06}, CSI is acquired using orthogonal pilot sequences, but, due to the shortage of orthogonal sequences, the same pilot sequences must be reused in neighboring cells, causing pilot contamination. This problem is considered as one of the major challenges in massive MIMO systems \cite{rusek13}. Mitigation of pilot contamination has been the focus of several works recently. These include \cite{gesbert13}, where it is utilized that the desired and interfering signals can be distinguished in the channel covariance matrices, as long as the angle-of-arrival spreads of desired and interfering signals do not overlap. A pilot sequence coordination scheme is proposed to help satisfying this condition. The work in \cite{ashikhmin12} utilizes coordination among base stations to share downlink messages. Each BS then performs linear combinations of messages intended for users applying the same pilot sequence. This is shown to eliminate interference when the number of base station antennas goes to infinity. A multi-cell precoding technique is used in \cite{jose11} with the objective of not only minimizing the mean squared error of the signals within the cell, but also minimizing the interference imposed to other cells. 


The survey of the related work indicates that the pilot contamination problem has been seen as an inter-cell problem that arises when the users associated with two neighboring cells use the same pilot sequence. An implicit assumption associated with it is that the pilot sequences of the users associated with the same cell are perfectly scheduled, such that no intra-cell pilot contamination occurs. These assumptions fall apart when one considers very dense, crowded scenarios as envisioned in 5G wireless scenarios \cite{METIS}. In such a setting, orthogonal scheduling of the users belonging to the same BS becomes infeasible, due to scheduling overhead. 

In this work, we consider such a crowd scenario, where the amount of users and their access behavior make it infeasible to schedule the transmissions. Instead users choose pilot sequences at random in an uncoordinated manner from a small pool shared by all users. Since the users are not coordinated, the pilot contamination problem can be cast as a \emph{random access} problem. We identify two features specific to massive MIMO: (1) asymptotic orthogonality between user channels; and (2) asymptotic invariance of the power received from a user over a short time interval. We use these features in order to formulate a \emph{pilot access protocol} using the framework of \emph{coded random access} \cite{Liva,SPV2012}. In such a framework, knowledge about the pilot applied by each individual user is not necessary a priori. This will be discussed in more detail in section \ref{sec:scheme}.

The difference from existing approaches for coded random access is that the proposed protocol combines decoding of the data in the uplink with estimation of the channel, which can be used for downlink transmission. Moreover, the massive MIMO property of a stable norm makes it possible to apply the protocol in fading channels, which is not possible with existing approaches to coded random access. Overall, the solution proposed in this paper is a radical departure from the usual treatment of the pilot contamination problem and introduces an important link to the area of random access protocols.

\section{System Model}\label{sec:sysmodel}
In this work we denote scalars in lower case, vectors in bold lower case and matrices in bold upper case. A superscript ``$T$'' denotes the transpose and a superscript ``$H$'' denotes the conjugate transpose.

\begin{figure}[t]
 \centering
 \includegraphics[width=1\columnwidth]{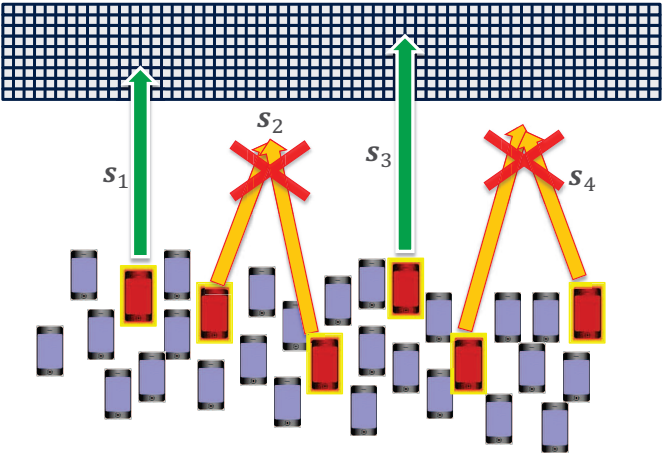}
 \caption{A single cell crowd scenario.}
 \label{fig:system}
\end{figure}

\begin{figure}[t]
 \centering
 \includegraphics[width=1\columnwidth]{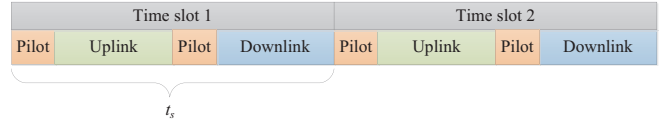}
 \caption{An example of a transmission schedule.}
 \label{fig:transschedule}
\end{figure}

We consider a random access system consisting of a single base station with $M$ antennas and $K$ users with a single antenna, where $M$ and $K$ are in the hundreds or thousands. see Fig.~\ref{fig:system}. Communication is performed on a time slotted basis, where each time slot consists of four phases; an uplink pilot phase, an uplink data phase, a downlink pilot phase and a downlink data phase, see Fig.~\ref{fig:transschedule}. The channel between the $k$'th user and the base station in the $n$'th time slot is denoted $\pmb{h}_{nk}=\left[h_{nk}(1)\hspace{0.1cm}h_{nk}(2) \hdots h_{nk}(M)\right]^T$, where $h_{nk}(i) \sim \mathcal{CN}(0,1)$ $\forall$ $i$. It is assumed that $\pmb{h}_{nk}$ $\forall$ $k$ are mutually orthogonal, which is justified by the range of $M$. Moreover, it is assumed that channel coefficients in different time slots are i.i.d, while the channel power, $\pmb{h}_{nk}^H\pmb{h}_{nk}=||\pmb{h}_{nk}||^2$ remains constant within a period of $\beta$ time slots. Note that the channel power varies due to path loss and shadowing effects, which causes it to vary much slower than the channel coefficients.

\begin{figure}[t]
 \centering
 \includegraphics[width=0.8\columnwidth]{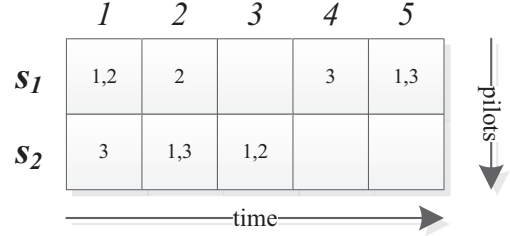}
 \caption{An example of a pilot schedule.}
 \label{fig:pilotschedule}
\end{figure}

In each time slot, each user is active with probability $p_a$. If a user is active, a random pilot sequence, $\pmb{s}_k=\left[s_k(1)\hspace{0.1cm}s_k(2) \hdots s_k(\tau)\right]^T$, is chosen among a set of size $\tau$ with mutually orthogonal pilot sequences. Note that multiple users may choose the same pilot sequence. See Fig.~\ref{fig:pilotschedule} for an example of a random pilot schedule with $\tau=2$ and $K=3$. By $\mathcal{A}_n$, we denote all active users in time slot $n$ and by $\mathcal{A}_n^j$, we denote the set of users applying $\pmb{s}_j$ in the $n$'th time slot. If $\pmb{Y}_{n}^{pu}$ denotes the uplink pilot signal received in time slot $n$, we have
\begin{align}\label{eq:pilots}
\pmb{Y}_{n}^{pu} = \sum_{j=1}^{\tau} \sum_{k\in\mathcal{A}_n^j} \pmb{h}_{nk} \pmb{s}_j^T + \pmb{Z}_{nj}^{pu},
\end{align}
\noindent where $\pmb{Z}_{nj}^{pu}$ is a matrix of i.i.d. Gaussian noise components, hence $\pmb{Z}_{nj}^{pu}(i,j) \sim \mathcal{CN}(0,\sigma_n^2)$ $\forall$ $i,j$. Any future instances of a vector $\pmb{z}$ or matrix $\pmb{Z}$, with different sub- or superscripts follow the same definition. All active users transmit a message of length $L$ in the uplink data phase. The message from the $k$'th user is denoted $\pmb{x}_k^u=\left[x_k^u(1)\hspace{0.1cm}x_k^u(2) \hdots x_k^u(L)\right]^T$. Denoting the received uplink signal in time slot $n$ as $\pmb{Y}_{n}^u$, we then have
\begin{align}\label{eq:udata}
\pmb{Y}_{n}^u = \sum_{k\in\mathcal{A}_n} \pmb{h}_{nk} {\pmb{x}_k^u}^T + \pmb{Z}_n^u.
\end{align}
In the downlink phase we rely on channel reciprocity, such that the uplink channel estimate is assumed to be a valid estimate of the downlink channel. The base station transmits a precoded downlink pilot sequence, such that the $k$'th user receives a downlink pilot signal, $\pmb{y}_{nk}^{pd}$, given by
\begin{align}\label{eq:pilots_down}
\pmb{y}_{nk}^{pd} = \pmb{h}_{nk}^T \pmb{w}_{nk} \pmb{s}_j^T + \pmb{z}_{nk}^{pd},
\end{align}
\noindent where $\pmb{w}_{nk}=\left[w_{nk}(1)\hspace{0.1cm}w_{nk}(2) \hdots w_{nk}(M)\right]^T$ is the precoding vector for user $k$ in the $n$'th time slot. The base station is able to schedule the downlink messages, $\pmb{x}_k^d=\left[x_k^d(1)\hspace{0.1cm}x_k^d(2) \hdots x_k^d(L)\right]^T$, such that the received signal in the downlink data phase is
\begin{align}\label{eq:ddata}
\pmb{y}_{nk}^d = \pmb{h}_{nk}^T \pmb{w}_{nk} {\pmb{x}_k^d}^T + \pmb{z}_{nk}^{d}.
\end{align}

%
\section{Pilot Access Protocol}\label{sec:scheme}
This section describes the proposed method of communication in the system described in section \ref{sec:sysmodel}. The main focus of this work is the uplink phase, however, a subsection is dedicated to describing the operation in the downlink phase.

\subsection{Uplink}
From the uplink pilot signals in \eqref{eq:pilots}, it is possible to estimate the channels between the users and the base station. However, since multiple users may apply the same pilot sequence, it is only possible to estimate a sum of the involved channels. The least squares estimate, $\pmb{\phi}_{nj}$, based on the pilot signal in time slot $n$ from users applying $\pmb{s}_j$ is found as
\begin{align}\label{eq:estimates}
\pmb{\phi}_{nj} &= ((\pmb{s}_j^H \pmb{s}_j)^{-1} \pmb{s}_j^H {\pmb{Y}_{n}^{pu}}^T)^T \notag \\
       &= \sum_{k\in\mathcal{A}_n^j} \pmb{h}_{nk} + \pmb{z}_{nj}^{pu'}.
\end{align}
\noindent where $\pmb{z}_{nj}^{pu'}$ is the impairment of the estimate caused by the noise, $\pmb{z}_{nj}^{pu}$. Any future instances of a vector $\pmb{z}$ with a prime follow the same definition.

The problem of interfering users applying the same, or a non-orthogonal, pilot sequence is often called \textit{pilot contamination}. If we proceed to detect the data in the uplink phase using a contaminated channel estimate, the result will be a summation of data messages. If $\pmb{\psi}_{nj}$ is the data estimate based on the channel estimate $\pmb{\phi}_{nj}$, we have
\begin{align}
\pmb{\psi}_{nj} &= ((\pmb{\phi}_{nj}^H \pmb{\phi}_{nj})^{-1} \pmb{\phi}_{nj}^H {\pmb{Y}_{n}^u})^T \notag \\
&= \sum_{k\in\mathcal{A}_n^j} \frac{||\pmb{h}_{nk}||^2}{||\pmb{h}_{nj}||^2} \pmb{x}_{k}^u + \pmb{z}_{n}^{u'}.
\end{align}
%
%
\noindent Hence, a pilot collision leads to a data collision. In our system, one way to deal with this problem is to carefully select $p_a$, such that the probability of having one and only one user applying a particular pilot sequence in a particular time slot is maximized. Hence, we have
\begin{align}\label{eq:max}
\underset{p_a}{\text{maximize}} & & \text{Pr}(\left\vert{\mathcal{A}_n^j}\right\vert = 1) \notag \\
\text{subject to} & & 0 \le p_a \le 1
\end{align}
\noindent This will maximize the number of non-contaminated channel estimates, and in turn maximize the number of successful data transmissions. This approach is reminiscent of the framed slotted ALOHA protocol for conventional random access. We consider this a reference scheme in this work and refer to it as ALOHA. Note that a random access, i.e. nonscheduled, scheme must be considered as a reference, due to the assumption of a crowded scenario, where scheduling is infeasible.

\begin{figure}[ht]
 \centering
 \includegraphics[width=1\columnwidth]{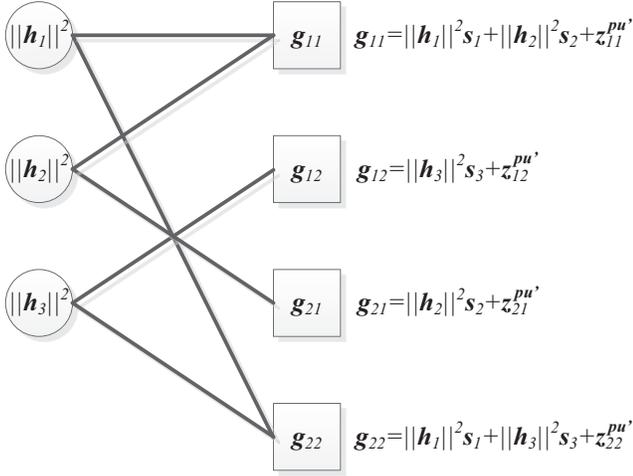}
 \caption{A graph representation of pilot collisions.}
 \label{fig:graphp}
\end{figure}

\begin{figure}[ht]
 \centering
 \includegraphics[width=1\columnwidth]{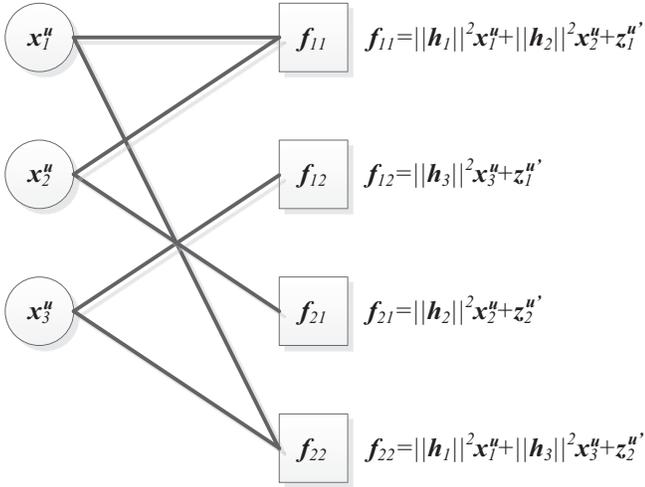}
 \caption{A graph representation of data collisions.}
 \label{fig:graphd}
\end{figure}

\begin{figure}[ht]
 \centering
 \includegraphics[width=1\columnwidth]{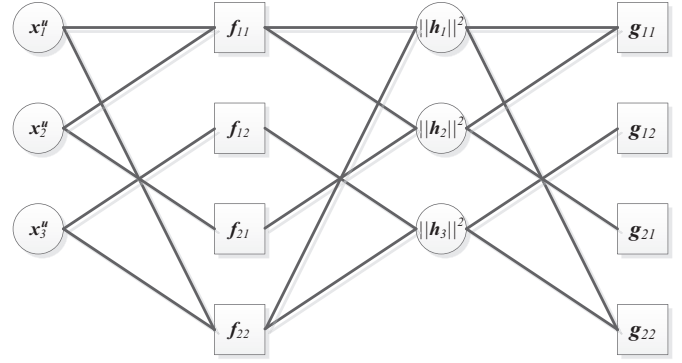}
 \caption{A graph representation of data and pilot collisions.}
 \label{fig:fullgraph}
\end{figure}

ALOHA has been state-of-the-art for many years within random access protocols, but recently a paradigm shift has started with the advent of coded random access \cite{Liva,SPV2012}. In this work, we view the problem of pilot contamination as a random access problem and apply newly developed tools in this area to solve the problem. Two features from the massive MIMO scenario are essential to our solution; near-orthogonality between user channels and near-stability of channel powers. Through signal processing techniques they allow us to resolve pilot collisions and thereby utilize otherwise wasted resources. The solution can be viewed as a two-stage processing approach:

\begin{enumerate}
\item \textbf{Matched filter:} The received uplink pilot and data signals, in \eqref{eq:pilots} and \eqref{eq:udata}, are processed using matched filters, which are constructed from the contaminated estimates in \eqref{eq:estimates}. More specifically, we multiply the received signals with $\pmb{\phi}_{nj}^H$ creating filtered signals, denoted $\pmb{f}_{nj}$ and $\pmb{g}_{nj}$ for data and pilots respectively. These signals contain linear combinations of the data and pilots transmitted by the users contributing to the contaminated estimate, $\pmb{\phi}_{nj}$, see \eqref{eq:filterf} and \eqref{eq:filterg}. The relationship between the variables we wish to estimate and the filtered signals can be viewed as a factor graph, see Fig.~\ref{fig:graphp} and Fig.~\ref{fig:graphd}.
\item \textbf{Successive interference cancellation (SIC):}
The coefficients of the linear combinations in \eqref{eq:filterf} and \eqref{eq:filterg} are the two-norms, $||\pmb{h}_{nk}||^2$, of the involved channels. In a massive MIMO system, these can be assumed slowly fading, contrary to the fast fading channel coefficients. Hence, successive interference cancellation can be applied on the filtered signals in order to reduce the linear combinations to data signals from individual users. This requires knowledge about the edges in the code graphs, i.e. what pilots have been applied by the individual users and in which time slots. This information is not available a priori at the base station. However, it can be embedded in the uplink data messages, such that when a data message has been recovered, the base station is informed about the pilot pattern chosen by the user. In practice, this could be realized by embedding the seed for a pseudo random number generator. Note, that graph knowledge is not necessary to initiate SIC, since a data message can be recovered immediately when one and only one user chose a particular pilot in a particular time slot. This provides the necessary graph information to proceed SIC using belief propagation. The overhead resulting from embedding graph information is considered negligible.
\end{enumerate}
\begin{align}
\pmb{f}_{nj} &= (\pmb{\phi}_{nj}^H {\pmb{Y}_{n}^u})^T \notag \\
       &= \sum_{k\in\mathcal{A}_n^j} ||\pmb{h}_{nk}||^2 \pmb{x}_k^u + \pmb{z}_{n}^{u'} \label{eq:filterf} \\
\pmb{g}_{nj} &= (\pmb{\phi}_{nj}^H {\pmb{Y}_{n}^{pu}})^T \notag \\
       &= \sum_{k\in\mathcal{A}_n^j} ||\pmb{h}_{nk}||^2 \pmb{s}_j + \pmb{z}_{nj}^{pu'} \label{eq:filterg}
\end{align}
The purpose of the matched filters is to transform the received signals from linear combinations with fast fading coefficients (the channel coefficients) into linear combinations with slowly fading coefficients (the norms). Note that the signals only remain linear combinations, when the channels are orthogonal, and that the coefficients are slowly fading only when the norms are stable. Both are fulfilled under the conditions given by a massive MIMO scenario. We can thus see the filtered signals, $\pmb{f}_{nj}$ and $\pmb{g}_{nj}$ $\forall$ $j$ and $n=1,2,...,\beta$, as a code on which iterative belief propagation can be performed. See Fig.~\ref{fig:fullgraph} for a graph showing the inter-dependencies between $\pmb{f}_{nj}$ and $\pmb{g}_{nj}$.
%

\textbf{Example:} Consider the simple example already introduced in Fig.~\ref{fig:pilotschedule}. We assume $\beta=2$, such that the resulting graphs after matched filtering are found in Fig.~\ref{fig:graphp} and Fig.~\ref{fig:graphd}. Note, that since the norms are assumed constant, we can omit the time index, such that $||\pmb{h}_{k}||^2=||\pmb{h}_{nk}||^2$ $\forall$ $n$. We then have
\begin{align}\label{eq:ex1}
\pmb{f}_{11} &= ||\pmb{h}_{1}||^2 \pmb{x}_1^u + ||\pmb{h}_{2}||^2 \pmb{x}_2^u + \pmb{z}_{11}^{u'}, \notag \\
\pmb{f}_{12} &= ||\pmb{h}_{3}||^2 \pmb{x}_3^u + \pmb{z}_{12}^{u'}, \notag \\
\pmb{f}_{21} &= ||\pmb{h}_{2}||^2 \pmb{x}_2^u + \pmb{z}_{21}^{u'}, \notag \\
\pmb{f}_{22} &= ||\pmb{h}_{1}||^2 \pmb{x}_1^u + ||\pmb{h}_{3}||^2 \pmb{x}_3^u + \pmb{z}_{22}^{u'}, \notag \\
\pmb{g}_{11} &= ||\pmb{h}_{1}||^2 \pmb{s}_1 + ||\pmb{h}_{2}||^2 \pmb{s}_1 + \pmb{z}_{11}^{pu'}, \notag \\
\pmb{g}_{12} &= ||\pmb{h}_{3}||^2 \pmb{s}_2 + \pmb{z}_{12}^{pu'}, \notag \\
\pmb{g}_{21} &= ||\pmb{h}_{2}||^2 \pmb{s}_1 + \pmb{z}_{21}^{pu'}, \notag \\
\pmb{g}_{22} &= ||\pmb{h}_{1}||^2 \pmb{s}_2 + ||\pmb{h}_{3}||^2 \pmb{s}_2 + \pmb{z}_{22}^{pu'}.
\end{align}
\noindent We introduce the variable $\pmb{c}$ which accounts for accumulated noise components and estimation errors. Note, that the magnitude of the elements in $\pmb{c}$ increases as processing progresses. This will be discussed in further detail in section \ref{sec:results}. 

Initially, we isolate the contribution from user $1$, giving us
\begin{align}\label{eq:ex2}
||\pmb{h}_{1}||^2 \pmb{x}_1^u + \pmb{c} = \pmb{f}_{11} - \pmb{f}_{21}.
\end{align}
\noindent Since the applied pilot sequence is known a priori by the base station, we can find the norm as
\begin{align}\label{eq:ex3}
||\pmb{h}_{1}||^2 + \pmb{c} = (\pmb{s}_1^H \pmb{s}_1)^{-1} \pmb{s}_1^H(\pmb{g}_{11} - \pmb{g}_{21}).
\end{align}
\noindent Finally, the estimate of the message from user $1$, $\hat{\pmb{x}}_1^u$, is
\begin{align}\label{eq:ex4}
\hat{\pmb{x}}_1^u &= ((\pmb{s}_1^H \pmb{s}_1)^{-1} \pmb{s}_1^H(\pmb{g}_{11} - \pmb{g}_{21}))^{-1}(\pmb{f}_{11} - \pmb{f}_{21}).
\end{align}
\noindent Similar operations can be performed for finding $\hat{\pmb{x}}_2^u$ and $\hat{\pmb{x}}_3^u$.

\subsection{Downlink}\label{sec:down}
In downlink we assume channel reciprocity, such that the user does not need to estimate each coefficient of $\pmb{h}_{nk}$, which would require a pilot signal for all $M$ antennas. Instead, we let the receiver estimate the concatenated ``channel'' consisting of both the downlink precoder, $\pmb{w}_{nk}$, and the actual channel. Denoting the concatenated channel, $q_{nk}$, we have
\begin{align}\label{eq:down}
q_{nk} &= \pmb{h}_{nk}^T \pmb{w}_{nk},
\end{align}
\noindent where $q_{nk}$ is estimated through \eqref{eq:pilots_down}.

In order to choose an appropriate precoder, the base station must have an estimate of the current channel. The coded operation applied in uplink does not guarantee that such an estimate is available. Uplink operation relies on SIC based only on knowledge of the norm. Hence, downlink transmission to a user is only possible if that user avoided collision during the previous uplink pilot phase, such that an uncontaminated channel estimate is available. This incurs a delay in downlink transmissions, whose magnitude is analyzed in section \ref{sec:analysis}.

\subsection{Analysis}\label{sec:analysis}
The performances of the reference scheme and the proposed scheme are tightly connected with the factor node degree distribution of the code graph. Here a factor degree, denoted as $d_{nj}$, refers to the number of users occupying the same resource block, i.e. applies the $j$'th pilot sequence in the $n$'th time slot. A user is active and applying pilot sequence $j$ with probability $p_a/\tau$, such that the degree probability distribution is
\begin{align}\label{eq:degree}
Pr(d_{nj}=d) &= \binom{K}{d}\left(\frac{p_a}{\tau}\right)^d\left(1-\frac{p_a}{\tau}\right)^{K-d}.
\end{align}
\noindent For the ALOHA scheme, we found that the optimal performance is achieved when the probability of having $d_{nj}=1$ is maximized. Differentiating $Pr(d_{nj}=1)$ with respect to $p_a$ and finding the roots satisfying our conditions, we get that $p_a=\frac{\tau}{K}$ maximizes the performance of the ALOHA scheme.


We can not use the same approach for optimizing the proposed scheme, since resource blocks with $d_{nj}>1$ may be useful. Instead we must seek a well performing degree distribution which favors the iterative belief propagation. Several works have studied this, e.g. in \cite{lt,sorensen}, however, in this work we can not freely tailor the degree distribution. We are limited to the binomial distribution as expressed in \eqref{eq:degree}, with only the freedom to choose a proper $p_a$. Similar limitations were considered in \cite{SPV2012} with focus on choosing an average degree, $\bar{d}$, which was optimized numerically. In our context, we have
\begin{align}\label{eq:bard}
\bar{d} = \frac{p_a K}{\tau}.
\end{align}
\noindent A numerical optimization of $\bar{d}$ and thereby in turn $p_a$ for a specific pair of $K$ and $\tau$ will be performed in section \ref{sec:results}.

As described in section \ref{sec:down}, downlink transmissions experience a delay due to lack of channel knowledge. We denote the delay for user $k$, $\Delta_k$. This delay is equal to the number of time slots until user $k$ is active and avoids a collision during the uplink pilot phase. Denoting the probability of a user being active and avoiding collision, $p_a^*$, we have 
\begin{align}\label{eq:delay}
p_a^* = p_a\left(1-\frac{p_a}{\tau}\right)^{K-1}.
\end{align}
\noindent The probability distribution of $\Delta_k$ follows the negative binomial and is therefore given by
\begin{align}\label{eq:delay_dist}
Pr(\Delta_k=\delta) = p_a^* (1-p_a^*)^{\delta-1}.
\end{align}
\noindent The expected value, $\mathrm{E}[\Delta_k]$, of the delay is then found as
\begin{align}\label{eq:delay_exp}
\mathrm{E}[\Delta_k] = \frac{(1-p_a^*)}{p_a^*}.
\end{align}
\noindent There is a natural tradeoff between optimizing $p_a$ for high uplink throughput and optimizing it for limiting the delay in the downlink phase. Such a joint optimization is outside the scope of this work. In the numerical evaluations in section \ref{sec:results}, we will solely be concerned with the uplink throughput.
\section{Numerical Results} \label{sec:results}
The proposed scheme is simulated and compared to framed slotted ALOHA in terms of uplink throughput and block error rate. Framed slotted ALOHA does not utilize SIC, but optimizes performance through a maximization of degree one nodes in the code graph, see \eqref{eq:max}. The proposed scheme is based on an assumption that the channel coefficients in different time slots are i.i.d, while their two-norms remain constant within a period of $\beta$ time slots. In the numerical evaluations, we will challenge these assumptions by simulating with fading channels. A rich scattering environment is assumed, such that $h_{nk}(m)$ can be modeled using Clarke's model \cite{clarke68}, hence
\begin{align}
h_{nk}(m) = \frac{1}{\sqrt{N_s}}\sum_{i=1}^{N_s} \mathrm{e}^{j2\pi f_d n t_s \cos\alpha_i+\phi_i},
\end{align}
\noindent where $N_s$ is the number of scatterers, $f_d$ is the maximum Doppler shift, $\alpha_i$ and $\phi_i$ is the angle of arrival and initial phase, respectively, of the wave from the $i$'th scatterer. Both $\alpha_i$ and $\phi_i$ are i.i.d. in the interval $[-\pi,\pi)$ and $f_d=\frac{v}{c}f_c$, where $v$ is the speed of the user, $c$ is the speed of light and $f_c$ is the carrier frequency. An overview of the simulation parameters is given in Table \ref{tab:params}. Note that $\beta=1.2K/\tau$ is chosen in order to ensure a 20\% surplus of resource blocks relative to $K$, such that the iterative belief propagation performs well. All simulation results are averages over $10,000$ iterations.

\begin{figure}[ht]
 \centering
 \includegraphics[width=1\columnwidth]{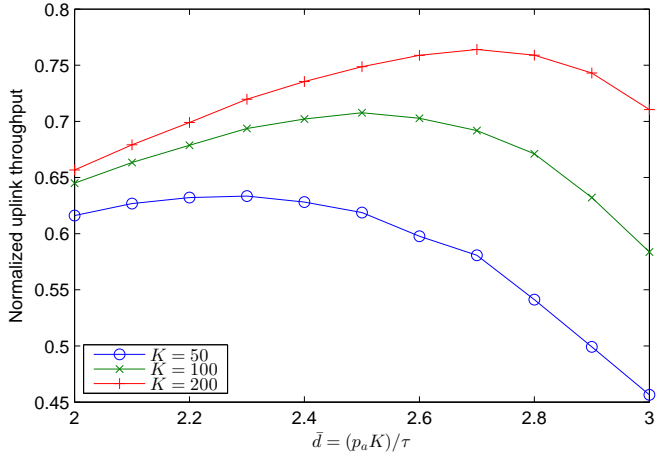}
 \caption{Throughput as a function of the average degree of a resource block.}
 \label{fig:tput_d}
\end{figure}

Initially, in Fig.~\ref{fig:tput_d} we present results for the normalized throughput of the proposed scheme as a function of the average degree of a resource block, which is directly related to the activation probability, as seen in \eqref{eq:bard}. We define normalized throughput as the total number of successfully decoded messages divided by $\beta\tau$, i.e. the total amount of resource blocks. It is evident that an average degree of approximately $2.5$ should be aimed for in the considered range of $K$, which is confirmed by the results from \cite{SPV2012}. All other simulations are performed using an average degree of $2.5$ regardless of $K$. Note that improved performance could be achieved by optimizing the activation probability to a particular value of $K$.

\begin{figure}[ht]
 \centering
 \includegraphics[width=1\columnwidth]{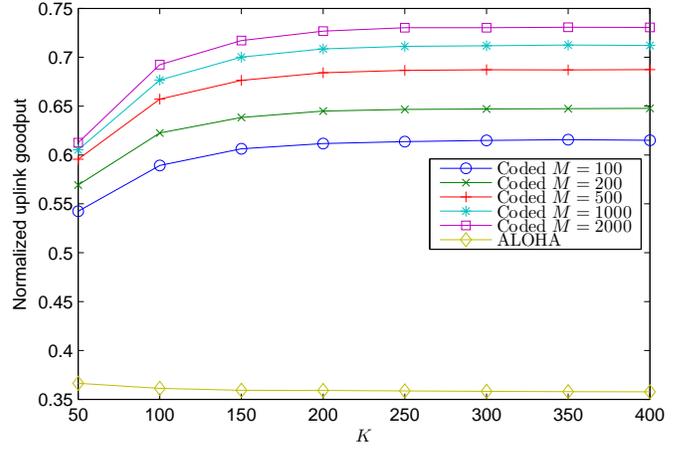}
 \caption{Throughput as a function of the number of users.}
 \label{fig:tput_N}
\end{figure}

Fig.~\ref{fig:tput_N} shows normalized goodput, i.e. throughput with erroneous messages discarded, as a function of the number of users accessing the base station. The proposed scheme clearly outperforms the conventional method of framed slotted ALOHA. The improvement increases with $K$, since the proposed scheme benefits from a larger number of messages to code across. An increase in $K$ can be viewed as an increase in the block length, which improves coding efficiency.

\begin{figure}[ht]
 \centering
 \includegraphics[width=1\columnwidth]{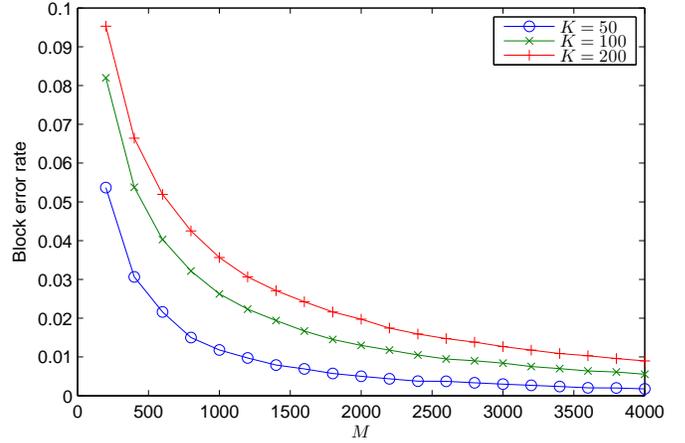}
 \caption{Block error rate as a function of the number of base station antennas.}
 \label{fig:BLER_M}
\end{figure}

The coding gain comes at the price of an increased error rate. As mentioned in section \ref{sec:scheme}, whenever SIC is performed, noise and estimation errors are accumulated, which may lead to errors. At higher $K$, it is more common to see high degrees in the code graph, even if the average degree remains constant. Moreover, SIC is performed across a larger time span, which leads to larger errors in the norm estimation. As a result, we experience an increased error rate for increasing $K$, which is illustrated in Fig.~\ref{fig:BLER_M}. It also shows that the error rate drops significantly, as the number of base station antennas increases. The reason is that the norm stabilizes for increasing $M$, making the assumption of a constant norm increasingly valid.

\begin{table}[ht]
\centering
	\caption{Simulation parameters}
	\label{tab:params}
\begin{tabular}{|c|c|c|} \hline
  Parameter & Value & Description \\ \hline \hline
  \rule{0pt}{3ex} $f_c$        & $1.8\hspace{0.1cm}$GHz  & Carrier frequency\\ \hline
  \rule{0pt}{3ex} $v$          & $3\hspace{0.1cm}$km/h   & User mobility\\ \hline
	\rule{0pt}{3ex} $N_s$        & $20$                    & Number of scatterers \\ \hline
	\rule{0pt}{3ex} $\sigma_n^2$ & $0.1$                   & Relative noise power \\ \hline
	\rule{0pt}{3ex} $\tau$       & $5\hspace{0.1cm}$bits   & Length and number of pilot sequences\\ \hline
	\rule{0pt}{3ex} $t_s$        & $0.01\hspace{0.1cm}$s   & Length of a time slot \\ \hline
	\rule{0pt}{3ex} $L$          & $1000\hspace{0.1cm}$bits& Length of uplink data messages\\ \hline
	\rule{0pt}{3ex} $\beta$      & $1.2K/\tau$             & Number of time slots\\ \hline
\end{tabular}
\end{table}

\section{Conclusions}\label{sec:conclusions}
We presented a solution for the pilot contamination problem in crowded scenarios, where users within a single cell must share a small set of pilot sequences. We view intra-cell pilot contamination as a random access problem and draw on newly developed ideas from this area of research. The massive MIMO setting provides two essential properties; near-orthogonality between user channels and near-stability of channel powers. These properties allow us to view a set of contaminated pilot signals as a graph code on which iterative belief propagation can be performed. The proposed solution proves highly efficient, comfortably outperforming the conventional ALOHA approach to random access. The price to pay is an increased error rate, due to accumulation of estimation errors in the belief propagation algorithm. However, this downside is shown to significantly diminish as the number of base station antennas increases.

\bibliographystyle{ieeetr}
\bibliography{bibliography}

\begin{thebibliography}{10}

\bibitem{marzetta06}
T.~Marzetta, ``How much training is required for multiuser mimo?,'' in {\em
  Signals, Systems and Computers, 2006. ACSSC '06. Fortieth Asilomar Conference
  on}, pp.~359--363, Oct 2006.

\bibitem{Petar5G}
F.~Boccardi, R.~Heath, A.~Lozano, T.~Marzetta, and P.~Popovski, ``Five
  disruptive technology directions for 5g,'' {\em Communications Magazine,
  IEEE}, vol.~52, pp.~74--80, February 2014.

\bibitem{rusek13}
F.~Rusek, D.~Persson, B.~K. Lau, E.~Larsson, T.~Marzetta, O.~Edfors, and
  F.~Tufvesson, ``Scaling up mimo: Opportunities and challenges with very large
  arrays,'' {\em Signal Processing Magazine, IEEE}, vol.~30, pp.~40--60, Jan
  2013.

\bibitem{gesbert13}
H.~Yin, D.~Gesbert, M.~Filippou, and Y.~Liu, ``A coordinated approach to
  channel estimation in large-scale multiple-antenna systems,'' {\em Selected
  Areas in Communications, IEEE Journal on}, vol.~31, pp.~264--273, February
  2013.

\bibitem{ashikhmin12}
A.~Ashikhmin and T.~Marzetta, ``Pilot contamination precoding in multi-cell
  large scale antenna systems,'' in {\em Information Theory Proceedings (ISIT),
  2012 IEEE International Symposium on}, pp.~1137--1141, July 2012.

\bibitem{jose11}
J.~Jose, A.~Ashikhmin, T.~Marzetta, and S.~Vishwanath, ``Pilot contamination
  and precoding in multi-cell tdd systems,'' {\em Wireless Communications, IEEE
  Transactions on}, vol.~10, pp.~2640--2651, August 2011.

\bibitem{METIS}
A.~Osseiran, F.~Boccardi, V.~Braun, K.~Kusume, P.~Marsch, M.~Maternia,
  O.~Queseth, M.~Schellmann, H.~Schotten, H.~Taoka, H.~Tullberg, M.~Uusitalo,
  B.~Timus, and M.~Fallgren, ``Scenarios for 5g mobile and wireless
  communications: the vision of the metis project,'' {\em Communications
  Magazine, IEEE}, vol.~52, pp.~26--35, May 2014.

\bibitem{Liva}
G.~Liva, ``{G}raph-{B}ased {A}nalysis and {O}ptimization of {C}ontention
  {R}esolution {D}iversity {S}lotted {ALOHA},'' {\em Communications, IEEE
  Transactions on}, vol.~59, pp.~477 --487, february 2011.

\bibitem{SPV2012}
C.~Stefanovic, P.~Popovski, and D.~Vukobratovic, ``{F}rameless {ALOHA} protocol
  for {W}ireless {Networks},'' {\em IEEE Comm. Letters}, vol.~16,
  pp.~2087--2090, Dec. 2012.

\bibitem{lt}
M.~Luby, ``{LT Codes},'' in {\em Foundations of Computer Science, 2002.
  Proceedings. The 43rd Annual IEEE Symposium on}, pp.~271--280, 2002.

\bibitem{sorensen}
J.~S{\o}rensen, P.~Popovski, and J.~{\O}stergaard, ``Design and analysis of lt
  codes with decreasing ripple size,'' {\em Communications, IEEE Transactions
  on}, vol.~60, pp.~3191 --3197, november 2012.

\bibitem{clarke68}
R.~Clarke, ``A statistical theory of mobile-radio reception,'' {\em Bell system
  technical journal}, vol.~47, no.~6, pp.~957--1000, 1968.

\end{thebibliography}

\end{document}